\documentclass{article}
\usepackage{amssymb}

\def\rset{\mathbb{R}}
\def\zset{\mathbb{Z}}

\def\ds{\displaystyle}

\def\pd#1#2{{\ds\partial #1\over\ds\partial #2}}

\newtheorem{thm}{Theorem}
\newtheorem{lm}{Lemma}

\title{Aubry-Mather Theory and Idempotent Eigenfunctions of Bellman
Operator\footnote{Comm. Contemporary Math. --- 1999. --- V. 1, N. 4. --- P. 517--533.}}
\author{A.N.~Sobolevski\u\i}
\date{Moscow State University\\
e-mail: \texttt{ansobol@idempan.phys.msu.su}}

\begin{document}

\maketitle

\begin{abstract}
We establish the connection between the Aubry-Mather theory of invariant
sets of a dynamical system described by a Lagrangian $L(t,x,v) = L_0(v) -
U(t,x)$ with periodic potential $U(t,x)$, on the one hand, and idempotent
spectral theory of the Bellman operator of the corresponding optimization
problem, on the other hand. This connection is applied to obtain a
uniqueness result for an eigenfunction of the Bellman operator in the case
of irrational rotation number.
\end{abstract}

\section{Introduction}

Consider a Lagrangian
$$
   L(t,x,v) = L_0(v) - U(t,x),
$$
where $t$,~$x \in \rset$ and the functions $L_0(v)$, $v \in \rset$, and
$U(t,x)$ satisfy the following conditions:
\begin{itemize}
\item[$(L_1)$] $L_0(p)$ is of class $C^1$; its derivative $L'_0(p)$ is
strictly growing: $L'_0(p_2) > L'_0(p_1)$ iff $p_2 > p_1$ and has a locally
Lipschitzian inverse.
\item[$(L_2)$] For any $N > 0$ there exists $P > 0$ such that if $p \in
\rset$, $|p| > P$, then $L_0(p) > N|p|$.
\item[$(L_3)$] $U(t,x)$ is of class $C^1$ and its derivative $\partial U/
\partial x$ is Lipschitzian.
\item[$(L_4)$] $U(t,x)$ is periodic: $U(t+1,x) = U(t,x+1) = U(t,x)$ for all
$(t,x) \in \rset^2$.
\end{itemize}

Let $x$, $y$, $s$, $t \in \rset$, $s < t$. We say that the set of all
absolutely continuous functions $\xi$:~$[s,t] \to \rset$ such that $\xi(s)
= y$, $\xi(t) = x$, and $L(\cdot, \xi(\cdot), \xi'(\cdot))$ is Lebesgue
integrable, is the set of \emph{admissible trajectories} (or
\emph{trajectories} for short) and denote it by $\Omega(s,y; t,x)$. The
functional defined on $\Omega(s,y; t,x)$ by the formula
$$
   \mathcal{L}(s,y;t,x)[\xi]
   = \int_s^t L(\tau, \xi(\tau), \xi'(\tau))\, d\tau
   = \int_s^t [L_0(\xi'(\tau)) - U(\tau,\xi(\tau))]\,d\tau
$$
is called the \emph{action functional} associated with the Lagrangian
$L(t,x,v)$. It follows from conditions~$(L_1)$--$(L_4)$ that this
functional is bounded from below. The infimum of $\mathcal{L}(s,y;t,x)$
over $\Omega(s,y;t,x)$ is called the \emph{value function} of the action
functional and is denoted by $L(s,y;t,x)$. It is well-known (see,
e.g.,~\cite{C}) that this infimum is attained, i.e., that there exists a
trajectory $\xi_0 \in \Omega(s,y;t,x)$ such that
$\mathcal{L}(s,y;t,x)[\xi_0] = L(s,y;t,x)$. This trajectory is called a
\emph{minimizer} of the functional $\mathcal{L}(s,y;t,x)$.

The Aubry-Mather theory~\cite{A,AlD,Mtop,MF} deals with a generalization of
the notion of minimizer to the case when $s$ and $t$ are allowed to be
infinite, $-\infty \le s < t \le \infty$. Since the potential $U(t,x)$ is
periodic in $t$, we may assume that $s$ and $t$ are integer or infinite and
consider any trajectory $\xi(t)$ only at integer moments of time: $x_n =
\xi(n)$. Any part of a minimizer is a minimizer itself; therefore the
minimization problem for the action functional can be discretized:
$$
   \sum_{k = s}^{t - 1} L(k, x_k; k + 1, x_{k + 1}) \to \min.
$$
It follows from periodicity of $U(t,x)$ in $t$ that $L(k,y;k + 1,x) =
L(0,y;1,x)$ for any $k \in \zset$; denote $L(0,y;1,x)$ by $L(y,x)$.

If $s = -\infty$ or $t = \infty$, then the above sum diverges. In this case
the definition of minimizer is modified in the following way: a sequence
$\{x_n\}$, $n \in \zset$, $s \le n \le t$, is said to be \emph{$L$-minimal}
if it has the following property: for all $n_1$, $n_2 \in \zset$, $s \le
n_1 < n_2 \le t$, and any set $\{y_n\}$, $n = n_1,n_1 + 1,\ldots,n_2$, such
that $y_{n_1} = x_{n_1}$ and $y_{n_2} = x_{n_2}$
$$
   \sum_{n = n_1}^{n_2 - 1} L(x_n,x_{n + 1}) \le
   \sum_{n = n_1}^{n_2 - 1} L(y_n,y_{n + 1}).
$$
This definition is due to Aubry~\cite{A,AlD}; similar concepts are known in
statistical mechanics (ground states in lattice models, see, e.g.,
\cite{S}) and differential geometry (geodesics of type $A$~\cite{Morse,H}).

It is clear that the definition of $L$-minimality does not change if
$L(y,x)$ is substituted by $L(y,x) + a(y - x)$ for any $a \in \rset$.
Consider an optimization problem of Bolza type
\begin{equation}
   s(x_{-n}) + \sum_{k = -n}^{-1} (L(x_k, x_{k + 1}) + a(x_k - x_{k + 1}))
   \to \hbox{min},
   \qquad x_0 = x.
\label{Bolza}
\end{equation}
It follows from Bellman's principle of optimality that for any solution
$\{x_{-n}, \allowbreak x_{-n + 1}, \ldots, x_0 = x\}$ of this problem the
following inclusion holds:
\begin{equation}
   x_{-k} \in \arg \min_{y \in \rset}\,
   ((B_a^{n - k}s)(y) + L(y,x_{-k + 1}) + a(y - x_{-k + 1})),
\label{inclusion}
\end{equation}
where operator $B_a$ is the so-called Bellman operator~\cite{KM,Yak}:
$$
   (B_a s)(x) = \min_{y \in \rset}\, (s(y) + L(y,x) + a(y - x)).
$$
It is clear that any solution of the problem~(\ref{Bolza}) satisfies the
condition of $L$-minimality.

Now suppose that a function $s^a(x)$ satisfies the following functional
equation:
\begin{equation}
   (B_a s^a)(x) = \min_{y \in \rset}\, (s^a(y) + L(y,x) + a(y - x))
   = s^a(x) + \lambda.
\label{functeq}
\end{equation}
Then any solution of the problem~(\ref{Bolza}) can be extended as $-n \to
-\infty$ using formula~(\ref{inclusion}) without violating the condition of
$L$-minimality. Thus to establish the existence of two-sided infinite
minimizers (corresponding to the case when $s = -\infty$, $t = \infty$) it
is sufficient to show that the initial point $x_0 = x$ can be chosen in
such a way that it is possible to find the reverse extension $\{x_1, x_2,
\ldots\}$ satisfying the condition of $L$-minimality. This program is
carried out in sections~3 and~4 after proving a number of technical lemmas
in section~2.

We stress that the functional equation~(\ref{functeq}), like many other
formulas in calculus of variations and optimization, can be treated as
linear over a special algebraic structure, an \emph{idempotent semiring}
$\rset_{\min}$~\cite{KM}. By definition, $\rset_{\min} = \{\rset \cup
\{\infty\}, \oplus, \odot\}$, where operations $\oplus$ and $\odot$ are
defined as follows: $a \oplus b = \min\{a,b\}$ (in particular, $a \oplus
\infty = \infty \oplus a = a$) and $a \odot b = a + b$ (in particular, $a
\odot \infty = \infty \odot a = \infty$).  It can easily be checked that
this structure satisfies axioms of an idempotent semiring with neutral
elements $\infty$ and $0$ with respect to operations $\oplus$ and $\odot$,
respectively~\cite{KM}. In this context, the Bellman operator assumes the
form of a linear integral operator:
$$
   (B_a s)(x) = \min_{y \in \rset}\, (s(y) + L(y,x) + a(y - x))
   = \int\limits_{y \in \rset}^\oplus s(y) \odot (L(y,x) + a(y - x))
$$
and the equation~(\ref{functeq}) defines its eigenfunction $s^a(x)$ and
eigenvalue $\lambda(a)$~\cite{KM,Yak}:
$$
   (B_a s)(x) = s(x) + \lambda(a) = (\lambda(a) \odot s)(x).
$$

This approach can be used further to re-establish the main results of
Aubry-Mather theory in an independent way; however, in this paper, we
assume a reverse viewpoint and apply Aubry-Mather theory to study of
idempotent eigenfunctions $s^a(x)$. One question that can be hard to
resolve in the general setting of idempotent analysis is whether an
eigenfunction corresponding to a given eigenvalue is unique. In section~5
we exploit minimality of Aubry-Mather set in the sense of topological
dynamics to prove that in the case of irrational rotation number the
eigenfunction is determined uniquely up to an additive constant. This
result is sharp in the sense that if the rotation number is rational, then
the solution may not be unique; a counterexample is constructed
in~\cite{JKM}.

This work was motivated by the preprint~\cite{JKM}, in which the problem
was stated in terms of periodic solution to periodically forced Burgers
equation rather than eigenfunction of Bellman operator. After the work was
finished~\cite{UMN}, the author learned that in the context of the forced
Burgers equation similar results, including uniqueness theorem, were
obtained independently by Weinan~E~\cite{E}.

The author is deeply grateful to Ya.~G.~Sinai for setting of the problem
and constant attention to this work and to Weinan~E for the opportunity to
read the unpublished version of his paper.

\section{Properties of the value function}

The following result is classical in calculus of variations (see,
e.g.,~\cite{C}):
\begin{lm}
Suppose $L_0(v)$ and $U(t,x)$ satisfy assumptions~$(L_1)$--$(L_4)$ and $x$,
$y$, $s$,~$t \in \rset$, $s < t$; then there exists a minimizer $\xi_0 \in
\Omega(s,y;t,x)$ such that $\mathcal{L}(s,y;t,x)[\xi_0] = L(s,y;t,x)$. If
$\xi_0 \in \Omega(s,y;t,x)$ is a minimizer, then $\xi_0$ is of class $C^1$
in $[s,t]$ and there exists a $C^1$ function $\lambda$:~$[s,t] \to \rset$
such that
\begin{equation}
   \lambda(\tau) = L'_0(\xi'_0(\tau)),\qquad
   \lambda'(\tau) = - \pd{U(\tau,\xi_0(\tau))}{\xi_0},\qquad
   s < \tau < t.
\label{Euler}
\end{equation}
For all $R$, $s$, $t \in \rset$ such that $R > 0$, $s < t$ and all $x$, $y
\in \rset$ such that $|x - y| \le R(t - s)$ there exists a positive
constant $C_1 = C_1(R,s,t)$ such that if $\xi_0 \in \Omega(s,y;t,x)$ is a
minimizer, then $|\xi'_0(\tau)| \le C_1$, $s \le \tau \le t$.
\label{l:Cesari}
\end{lm}

\begin{lm}
The function $L(s,y;t,x)$ is diagonally periodic:
$$
   L(s,y+1;t,x+1) = L(s+1,y;t+1,x) = L(s,y;t,x).
$$
\label{l:diagper}
\end{lm}
This lemma follows from periodicity of $U(t,x)$ in $t$ and $x$.

\begin{lm}
For all $r$, $s$, $t$, $x$, $y \in \rset$, $s < r < t$,
\begin{equation}
   L(s,y;t,x) = \min_{z \in \rset}\,(L(s,y;r,z) + L(r,z;t,x)).
\label{Bellman}
\end{equation}
\label{l:Bellman}
\end{lm}
This is a variant of the well-known Bellman's principle of optimality.

\begin{lm}
The function $L(s,y;t,x)$ satisfies inequalities
\begin{equation}
   -M \le {1\over t-s}L(s,y;t,x) - L_0\left({x-y\over t-s}\right) \le -m,
\label{updown}
\end{equation}
where $m = \min_{(t,x) \in \rset^2}\,U(t,x)$, $M = \max_{(t,x) \in
\rset^2}\,U(t,x)$.
\label{l:updown}
\end{lm}
\textsc{Proof.} Consider the functional
$$
   \widehat{\mathcal{L}}(s,y;t,x)[\xi]
   = \int_s^t L_0(\xi'(\tau))\,d\tau.
$$
The trajectory $\widehat\xi(\tau) = x(\tau-s)/\mbox{$(t-s)$} +
y(t-\tau)/(t-s)$, $s \le \tau \le t$, is a minimizer of this functional in
the set $\Omega(s,y;t,x)$ and
$$
   \widehat{\mathcal{L}}(s,y;t,x)[\widehat\xi] =
   (t - s) L_0\left({x - y\over t - s}\right).
$$

Let $\xi_0$ be a minimizer of the functional $\mathcal{L}(s,y;t,x)$ in
the class $\Omega(s,y;t,x)$, i.e., $\mathcal{L}(s,y;t,x)[\xi_0] =
L(s,y;t,x)$. We have
$$
\begin{array}{l}
   \ds L(s,y;t,x) - (t-s)L_0\left({x-y\over t-s}\right)
   = \mathcal{L}(s,y;t,x)[\xi_0]
   - \widehat{\mathcal{L}}(s,y;t,x)[\widehat\xi]) = \\
   \ds\qquad= -\int_s^t U(\tau,\xi_0(\tau))\,d\tau
   + (\widehat{\mathcal{L}}(s,y;t,x)[\xi_0]
   - \widehat{\mathcal{L}}(s,y;t,x)[\widehat\xi]) \ge \\
   \ds\qquad\ge (t-s)\left(-\max_{(t,x) \in \rset^2} U(t,x)\right)
\end{array}
$$
and
$$
\begin{array}{l}
   \ds L(s,y;t,x) - (t-s)L_0\left({x-y\over t-s}\right)
   = \mathcal{L}(s,y;t,x)[\xi_0]
   - \widehat{\mathcal{L}}(s,y;t,x)[\widehat\xi] \le \\
   \ds \qquad\le \mathcal{L}(s,y;t,x)[\widehat\xi]
   - \widehat{\mathcal{L}}(s,y;t,x)[\widehat\xi] = \\
   \ds \qquad = -\int_s^t U(\tau,\widehat\xi(\tau))\,d\tau
   \le (t-s)\left(-\min_{(t,x) \in \rset^2} U(t,x)\right).
\end{array}
$$
Multiplying these inequalities by $(t-s)^{-1}$, we obtain~(\ref{updown}).
\hfill$\square$

\begin{lm}
The function $L(s,y;t,x)$ is locally Lipschitzian:  if $R$, $x_1$, $x_2$,
$y_1$, $y_2$, $s$, $t \in \rset$, $s < t$, $R > 0$, $|x_i - y_j| \le
R(t-s)$, $i$,~$j = 1,2$, then there exists $C = C(R,s,t) \in \rset$,
$C > 0$, such that
\begin{equation}
   |L(s,y_1;t,x_1) - L(s,y_2;t,x_2)| \le C(|y_1 - y_2| + |x_1 - x_2|).
\label{Lip}
\end{equation}
\label{l:Lip}
\end{lm}
\smallskip
\par\noindent\textsc{Proof.} It is sufficient to prove the inequality
\begin{equation}
   |L(s,y;t,x_1) - L(s,y;t,x_2)| \le C|x_1 - x_2|,
\label{temp1}
\end{equation}
where $|x_i - y| \le R(t-s)$, $i = 1,2$. The inequality for the first pair
of arguments of the function $L$ is proved similarly; combining these two
inequalities, we obtain~(\ref{Lip}).

Let $\xi_0 \in \Omega(s,y;t,x_1)$ be a minimizer of the functional
$\mathcal{L}(s,y;t,x_1)$. Consider the trajectory $\xi(\tau) = \xi_0(\tau)
+ (x_2 - x_1)(\tau - s) / (t-s)$, $s \le \tau \le t$. It is clear that $\xi
\in \Omega(s,y;t,x_2)$.  We have
$$
\begin{array}{l}
   \ds L(s,y;t,x_2) - L(s,y;t,x_1)
   \le \mathcal{L}(s,y;t,x_2)[\xi] - \mathcal{L}(s,y;t,x_1)[\xi_0] = \\
   \ds\qquad = \int_s^t \left[L_0\left(\xi'_0(\tau)
   + {x_2 - x_1\over t-s}\right) - L_0(\xi'_0(\tau))\right]\,d\tau - \\
   \ds\qquad\quad- \int_s^t \left[U\left(\tau, \xi_0(\tau)
   + {x_2 - x_1\over t-s}(\tau - s)\right)
   - U(\tau,\xi_0(\tau))\right]\,d\tau.
\end{array}
$$
The functions $U(t,x)$ and $L_0(v)$ are of class $C^1$; in addition, it
follows from~$(L_3)$, $(L_4)$ that $|\partial U(t,x)/\partial x| \le M_1$
for some positive $M_1 \in \rset$. Hence,
$$
\begin{array}{l}
   \ds L(s,y;t,x_2) - L(s,y;t,x_1) \le \\
   \ds\qquad\le {\ds|x_2 - x_1|\over \ds t - s}(t - s)
   \max_{\tau \in [s,t], |\theta| \le 1}
   \left|L'_0\left(\xi'_0(\tau) + \theta{|x_2 - x_1|\over t-s}\right)\right|
   + \\
   \ds\qquad\quad + {\ds|x_2 - x_1|\over \ds t-s}{(t-s)^2\over 2}M_1.
\end{array}
$$
By lemma~\ref{l:Cesari}, it follows that $|\xi_0'(\tau)| \le C_1 =
C_1(R,s,t)$, $s \le \tau \le t$. Also, by~$(L_1)$, $L'_0(v)$ is finite for
all $v$; thus for some $C = C(M_1,R,s,t) \in \rset$, $C > 0$,
$$
   L(s,y;t,x_2) - L(s,y;t,x_1) \le C|x_2 - x_1|.
$$
Similarly, $L(s,y;t,x_2) - L(s,y;t,x_1) \ge -C|x_2 - x_1|$. This completes
the proof of inequality~(\ref{temp1}). \hfill$\square$

\begin{lm}
Let $x_1$, $x_2$, $y_1$, $y_2$, $s$, $r$, $t \in \rset$, $s < r < t$;
suppose there exists $z_0 \in \rset$ such that for all $z \in \rset$
$$
   L(s,y_i;r,z) + L(r,z;t,x_i) \ge L(s,y_i;r,z_0) + L(r,z_0;t,x_i),\quad
   i = 1,2.
$$
Then either $x_1 = x_2$ and $y_1 = y_2$ or $(x_1 - x_2)(y_1 - y_2) < 0$.
\label{l:Bangert}
\end{lm}
\textsc{Proof.} For $x_1 = x_2$ and $y_1 = y_2$ there is nothing to prove.
Assume $x_1 \neq x_2$ or $y_1 \neq y_2$. Let $\xi_i^< \in
\Omega(s,y_i;r,z_0)$ and $\xi_i^> \in \Omega(r,z_0;t,x_i)$ be minimizers of
functionals $\mathcal{L}(s,y_i;r,z_0)$ and $\mathcal{L}(r,z_0;t,x_i)$, $i =
1,2$. It follows from the condition of the lemma and Bellman's principle of
optimality (lemma~\ref{l:Bellman}) that for $i = 1,2$
$$
   L(s,y_i;r,z_0) + L(r,z_0;t,x_i) = L(s,y_i;t,x_i).
$$
Hence for $i = 1,2$ the trajectories
$$
   \xi_i(\tau) = \left\{
   \begin{array}{ll}
      \xi_i^<(\tau), &s \le \tau \le r \\
      \xi_i^>(\tau), &r \le \tau \le t \\
   \end{array}\right.
   \quad(\xi_i \in \Omega(s,y_i;t,x_i))
$$
are minimizers of $\mathcal{L}(s,y_i;t,x_i)$. Using
lemma~\ref{l:Cesari}, we see that the trajectories $\xi_i$, $i =
1,2$, are of class $C^1$. This means that $(\xi^<_i)'(r-0) =
(\xi^>_i)'(r+0) = \xi'_i(r)$, $i = 1,2$.

We claim that $\xi'_1(r) \neq \xi'_2(r)$. Indeed, let $\lambda_i$, $i =
1,2$, be the $C^1$ functions corresponding to trajectories $\xi_i$ by
lemma~\ref{l:Cesari}. Assume that $\xi'_1(r) = \xi'_2(r)$; then, by
continuity of $L'_0(\cdot)$, it follows from~(\ref{Euler}) that
$\lambda_1(r) = \lambda_2(r)$. Consider the Cauchy problem for the system
of ordinary differential equations~(\ref{Euler}) on the interval $[s,t]$
with initial data $\xi_0(r) = \xi_1(r) = \xi_2(r)$ and $\lambda(r) =
\lambda_1(r) = \lambda_2(r)$. It follows from condition $(L_3)$ that the
function $\partial U(t,x)/\partial x$ is Lipschitzian and its absolute
value is bounded by the constant $M_1 > 0$; thus it follows from the second
equation~(\ref{Euler}) that $|\lambda(\tau) - \lambda(r)| \le M_1(t - s)$
for all $\tau \in [s,t]$. Hence it follows from condition $(L_1)$ that the
system
$$
   \xi'_0(\tau) = (L'_0)^{-1}(\lambda(\tau)), \qquad
   \lambda'(\tau) = - \pd{U(\tau, \xi_0(\tau))}{\xi_0}, \qquad
   s < \tau < t,
$$
which is equivalent to~(\ref{Euler}), satisfies the conditions of the
uniqueness theorem. Thus $\xi_1(\tau) = \xi_2(\tau)$, $s \le \tau \le t$,
and in particular $x_1 = x_2$ and $y_1 = y_2$. This contradiction proves
that $\xi'_1(r) \neq \xi'_2(r)$.

Now we prove that there exists $\delta > 0$ such that $s < r - \delta$, $r
+ \delta < t$, and if $\tau \in (r-\delta, r+\delta)$, $\tau \neq r$, then
$\xi_1(\tau) \neq \xi_2(\tau)$. Indeed, let a sequence $\{t_n\}$ be such
that $s < t_n < t$, $t_n \neq r$, $\xi_1(t_n) = \xi_2(t_n)$ for all $n$,
and $\lim_{n \to \infty} t_n = r$. This implies that
$$
   \xi'_1(r)
   = \lim_{n \to \infty}{\xi_1(t_n) - \xi_1(r)\over t_n - r}
   = \lim_{n \to \infty}{\xi_2(t_n) - \xi_2(r)\over t_n - r}
   = \xi'_2(r),
$$
a contradiction.

Now assume that $(x_1 - x_2)(y_1 - y_2) \ge 0$; to be precise, let $x_1 \le
x_2$ and $y_1 \le y_2$. For $i = 1,2$ define trajectories $\bar\xi_i \in
\Omega(s,y_i;t,x_i)$ by $\bar\xi_1(\tau) = \min\{\xi_1(\tau),
\xi_2(\tau)\}$ and $\bar\xi_2(\tau) = \max\{\xi_1(\tau), \xi_2(\tau)\}$. By
the above each of $\bar\xi_1$ and $\bar\xi_2$ coincides with $\xi_1$ and
$\xi_2$ on a finite number of intervals of finite length. Let us check
that the trajectories $\bar\xi_i$, $i = 1,2$, are minimizers. Indeed, we
have
$$
   \mathcal{L}(s,y_i;t,x_i)[\bar\xi_i]
   \ge \mathcal{L}(s,y_i;t,x_i)[\xi_i]
   = L(s,y_i;t,x_i),
   \qquad i = 1,2,
$$
and
$$
\begin{array}{l}
   \mathcal{L}(s,y_1;t,x_1)[\bar\xi_1]
   + \mathcal{L}(s,y_2;t,x_2)[\bar\xi_2] = \\
   \quad = \mathcal{L}(s,y_1;t,x_1)[\xi_1]
   + \mathcal{L}(s,y_2;t,x_2)[\xi_2] = \\
   \quad = L(s,y_1;t,x_1) + L(s,y_2;t,x_2).
\end{array}
$$
Therefore $\mathcal{L}(s,y_i;t,x_i)[\bar\xi_i] = L(s,y_i;t,x_i)$, $i =
1,2$.

By lemma~\ref{l:Cesari}, the trajectories $\bar\xi_i$, $i = 1,2$, are
of class $C^1$. This implies that $\bar\xi'_i(r-0) = \bar\xi'_i(r+0)$, $i =
1,2$, where all one-sided derivatives exist. But this means that $\xi'_1(r
- 0) = \xi'_2(r + 0)$ and $\xi'_2(r - 0) = \xi'_1(r + 0)$, since $r$ is the
only intersection point of $\xi_1$ and $\xi_2$ in the interval $(r -
\delta, r + \delta)$. Thus $\xi'_1(r) = \xi'_2(r)$. This contradiction
concludes the proof. \hfill$\square$
\medskip
\par\noindent\textbf{Corollary} \emph{Let $x_1$, $x_2$, $y_1$, $y_2$, $s$,
$t \in \rset$, $x_1 < x_2$, $y_1 < y_2$, and $s < t$. Then}
\begin{equation}
   L(s,y_1;t,x_1) + L(s,y_2;t,x_2)
   < L(s,y_2;t,x_1) + L(s,y_1;t,x_2).
\label{twist}
\end{equation}
\medskip
\par\noindent\textsc{Proof.} Let $\xi_1 \in \Omega(s,y_1;t,x_2)$ and $\xi_2
\in \Omega(s,y_2;t,x_1)$ be minimizers of functionals
$\mathcal{L}(s,y_1;t,x_2)$ and $\mathcal{L}(s,y_2;t,x_1)$, respectively.
Note that $\xi_1(s) < \xi_2(s)$ and $\xi_1(t) > \xi_2(t)$; hence there
exist $r$ and $z_0$ such that $s < r < t$, $\xi_1(r) = \xi_2(r) = z_0$.
Therefore $L(s,y_1;r,z_0) + L(r,z_0;t,x_2) = L(s,y_1;t,x_2)$,
$L(s,y_2;r,z_0) + L(r,z_0;t,x_1) = L(s,y_2;t,x_1)$ and by
lemma~\ref{l:Bellman} $ L(s,y_1;r,z_0) + L(r,z_0;t,x_1) \ge
L(s,y_1;t,x_1)$, $L(s,y_2;r,z_0) + L(r,z_0;t,x_2) \ge L(s,y_2;t,x_2)$. This
implies that $L(s,y_1;t,x_1) + L(s,y_2;t,x_2) \le L(s,y_1;t,x_2) +
L(s,y_2;t,x_1)$.  Assume that this inequality is actually an equality.
This assumption means that for all $z \in \rset$
$$
\begin{array}{c}
   \ds L(s,y_1;t,x_1)
   = L(s,y_1;r,z_0) + L(r,z_0;t,x_1)
   \le L(s,y_1;r,z) + L(r,z;t,x_1), \\
   \ds L(s,y_2;t,x_2)
   = L(s,y_2;r,z_0) + L(r,z_0;t,x_2)
   \le L(s,y_2;r,z) + L(r,z;t,x_2).
\end{array}
$$
It follows from the previous lemma that $(x_1 - x_2)(y_1 - y_2) < 0$.
This contradiction concludes the proof. \hfill$\square$

\section{Existence of an eigenfunction of the Bellman operator}

Let $a \in \rset$, $s_0(x)$ be a continuous periodic function. We define
$$
   s_n(x;a,s_0) = (B_a^ns_0)(x) =
   \min_{y \in \rset}\,(s_0(y) + L(0,y;n,x) + a(y - x)).
$$
As a function of $x$, $s_n(x;a,s_0)$ is bounded from below. Further, it follows from
diagonal periodicity of $L(s,y;t,x)$ (lemma~\ref{l:diagper}) that it is
periodic; thus $s_n(x;a,s_0) = \min_{k \in \zset}\,s_n(x - k; a,s_0)$.
Using this formula, lemma~\ref{l:diagper}, and periodicity of $s_0(y)$, we
get
$$
   s_n(x;a,s_0) = \min_{k \in \zset} \min_{y \in \rset}\,
   (s_0(y) + L(0,y + k;n,x) + a(y + k - x)).
$$
Thus for any $n > 0$ and $y$, $x \in \rset$ the quantity $L(0,y+k;n,x) +
a(y + k - x)$ is bounded from below as a function of $k \in \zset$. We
denote
\begin{equation}
   L^a_n(y,x) = \min_{k \in \zset}\,(L(0,y + k;n,x) + a(y + k - x)),
\label{defLan}
\end{equation}
where minimum is attained since $L(s,y;t,x)$ is continuous and grows
arbitrarily large as $|x - y| \to \infty$; hence,
\begin{equation}
   s_n(x;a,s_0) = \min_{y \in \rset}\,(s_0(y) + L^a_n(y,x)).
\label{defsn}
\end{equation}
\begin{lm}
For any $n = 1,2,\ldots,$ the function $L^a_n(y,x)$ is periodic in both
variables and Lipschitzian in $x$; the Lipschitz constant $C^*(a)$ does not
depend on $n$ and $y \in \rset$. For any $n_1 > 0$, $n_2 > 0$
\begin{equation}
   L^a_{n_1 + n_2}(y,x)
   = \min_{z \in \rset}\,(L^a_{n_1}(y,z) + L^a_{n_2}(z,x)).
\label{Lan1n2}
\end{equation}
\label{l:LanLip}
\end{lm}
\textsc{Proof.} Using the definition~(\ref{defLan}) and
lemma~\ref{l:diagper}, we obtain $L^a_n(y + l,x + m) = L^a_n(y,x)$ for any
integer $l$, $m$; thus $L^a_n(y,x)$ is periodic. Combining~(\ref{defLan})
for $n = n_1 + n_2$ with Bellman's principle of optimality (\ref{Bellman})
for $s = 0$, $r = n_1$, and $t = n_1 + n_2$ and using
lemma~\ref{l:diagper}, we get
$$
   L^a_{n_1+n_2}(y,x)
   = \min_{z \in \rset}\,(L^a_{n_1}(y,z) + L(0,z;n_2,x) + a(z-x)).
$$
Using periodicity of $L^a_{n_1}(y,z)$ in $z$, we obtain~(\ref{Lan1n2}).

Let us prove that $L^a_1(y,x)$ is Lipschitzian. Since $L^a_1(y,x)$ is a
pointwise minimum of a family of continuous functions, it is upper semicontinuous.
Denote by $N$ the maximum value of $L^a_1(y,x)$ on the set $Q = \{\, (y,x)
\mid 0 \le y \le 1, 0 \le x \le 1 \,\}$. Clearly, there exists $R > 0$ such
that $L_0(x - y - k) + a(y + k - x) > N + \max_{(t,x) \in \rset^2}\,U(t,x)$
for all $(y,x) \in Q$ whenever $|k| > R$. Thus it follows from
lemma~\ref{l:updown} that for any $(y,x) \in Q$
$$
   L^a_1(y,x) = \min_{k \in \zset, |k| \le R}\,
   (L(0,y + k;1,x) + a(y + k - x)).
$$
On the other hand, using lemma~\ref{l:Lip}, we see that all functions
$L(0,y + k;1,x) + a(y + k - x)$, $|k| < R$, are Lipschitzian on $Q$ with
the constant $C^*(a) = C(R + 1,0,1) + |a|$. Thus $L^a_1(y,x)$ is
Lipschitzian on $Q$ (and, by periodicity, on the whole $\rset^2$) with the
same constant.

Now it follows from~(\ref{Lan1n2}) that for any $n \ge 2$
$$
   L^a_n(y,x) = \min_{z \in \rset}\,(L^a_{n-1}(y,z) + L^a_1(z,x)),
$$
that is for any $y \in \rset$ $L^a_n(y,x)$ as a function of $x$ is a
pointwise minimum of a family of functions having the same Lipschitz
constant $C^*(a)$. Thus $L^a_n(y,x)$ is Lipschitz continuous in $x$ with
the constant $C^*(a)$ for all $n = 1,2,\ldots$
\hfill$\square$
\medskip
\par\noindent\textbf{Corollary} \emph{For any $n = 1,2,\ldots$ the function
$s_n(x;a,s_0)$ defined in~(\ref{defsn}) is Lipschitzian with the constant
$C^*(a)$; it satisfies}
\begin{equation}
   \max_{x \in \rset}\,s_n(x;a,s_0)
   - \min_{x \in \rset}\,s_n(x;a,s_0)
   \le C^*(a)
\label{oscsn}
\end{equation}
\emph{and for all $m \in \zset$, $m \ge n$,}
\begin{equation}
   s_m(x;a,s_0)
   = \min_{y \in \rset}\,(s_{m-n}(y;a,s_0) + L^a_n(y,x)).
\label{smsn}
\end{equation}
\medskip
\par\noindent\textsc{Proof.} From~(\ref{defsn}) it follows that
$s_n(x;a,s_0)$ is a pointwise minimum of a family of Lipschitzian functions
having the Lipschitz constant $C^*(a)$; hence it is Lipschitzian with the
same constant. Taking into account periodicity of $s_n(x;a,s_0)$, we
obtain~(\ref{oscsn}).  Equation~(\ref{smsn}) follows from~(\ref{Lan1n2})
for $n_1 = m - n$, $n_2 = n$. \hfill$\square$

Denote
\begin{equation}
   \lambda_n(a)
   = \min_{(y,x) \in \rset^2}\,{1\over n}L^a_n(y,x)
   = \min_{(y,x) \in \rset^2}\,
   \left({1\over n}L(0,y;n,x) + a{y-x\over n}\right).
\label{defHn}
\end{equation}
\begin{lm}
For any $a \in \rset$ there exists $\lambda(a) \in \rset$ such that
\begin{equation}
   |\lambda_n(a) - \lambda(a)| \le {C^*(a)\over n},
\label{HnHa}
\end{equation}
where $C^*(a)$ is the Lipschitz constant of the function $L^a_n(y,x)$. The
function $a \mapsto \lambda(a)$ is concave and satisfies inequalities
\begin{equation}
   H_0(a) + m \le -\lambda(a) \le H_0(a) + M,
\label{estHH0rep}
\end{equation}
where $m$ and $M$ are defined in lemma~\ref{l:updown} and $H_0(a)$ is the
Legendre transform of $L_0(v)$.
\label{l:Ha}
\end{lm}
\textsc{Proof.} Let $s_0(x) = 0$, $s_n(x) = s_n(x;a,s_0)$.
From~(\ref{smsn}), for any integer $n$, $n_0$, $0 < n_0 < n$,
\begin{equation}
   s_{n}(x) = \min_{y \in \rset}\,(s_{n - n_0}(y) + L^a_{n_0}(y,x)).
\label{temp3}
\end{equation}
We see that $\min_{x \in \rset}\,s_n(x) = \min_{(y,x) \in \rset^2}\,
L^a_n(y,x) = n \lambda_n(a)$ for any $n = 1,2,\ldots$ It follows from the
corollary to lemma~\ref{l:LanLip} that $n\lambda_n(a) \le s_n(x) \le
n\lambda_n(a) + C^*(a)$. Combining these inequalities with~(\ref{temp3})
and~(\ref{defHn}), we obtain
$$
\begin{array}{l}
   n\lambda_n(a)
   \le s_n(x)
   \le (n - n_0)\lambda_{n - n_0}(a) + C^*(a) + n_0 \lambda_{n_0}(a),\\
   (n - n_0)\lambda_{n - n_0}(a) + n_0 \lambda_{n_0}(a)
   \le s_n(x)
   \le n\lambda_n(a) + C^*(a).
\end{array}
$$
Hence,
$$
   |n\lambda_n(a) - n_0\lambda_{n_0}(a) - (n - n_0)\lambda_{n - n_0}(a)|
   \le C^*(a).
$$
There exist integer $p \ge 1$ and $0 \le q < n$ such that $n = pn_0 + q$.
By induction over $p$ it is easily checked that $|n \lambda_n(a) - pn_0
\lambda_{n_0}(a) - q\lambda_q(a)| \le pC^*(a)$, that is
\begin{equation}
   \left| \lambda_n(a) - \lambda_{n_0}(a) + {q\over n}(\lambda_{n_0}(a)
   - \lambda_q(a))\right| \le {p\over n}C^*(a) \le {1\over n_0}C^*(a).
\label{xxx}
\end{equation}
Let $\varepsilon > 0$ be arbitrarily small, $n_0$ be so large that the
right-hand side of this inequality is less than $\varepsilon$, and $N$ be
such that $(1/N)\max_{1 \le q \le n_0}\, |q(\lambda_{n_0}(a) -
\lambda_q(a)| < \varepsilon$; then $|\lambda_{n_1}(a) - \lambda_{n_2}(a)| <
4\varepsilon$ for all $n_1$, $n_2 \ge N$. Thus $\{\lambda_n(a)\}$, $n =
1,2,\ldots,$ is a Cauchy sequence. Denote its limit by $\lambda(a)$; then
we get~(\ref{HnHa}) from~(\ref{xxx}) in the limit $n \to \infty$.

Suppose $a_1$,~$a_2 \in \rset$, $0 < \alpha, \beta < 1$, $\alpha + \beta =
1$.  Using definition of $\lambda_n(a)$~(\ref{defHn}), we obtain
$$
\begin{array}{l}
   \ds \lambda_n(\alpha a_1 + \beta a_2)
   = {1\over n} \min_{(y,x) \in \rset^2}\,
   (L(0,y;n,x) + (\alpha a_1 + \beta a_2)(y - x)) = \\
   \ds\quad = {1\over n} \min_{(y,x) \in \rset^2}\,
   [\alpha(L(0,y;n,x) + a_1(y - x)) + \beta(L(0,y;n,x) + a(y - x))) \ge \\
   \ds\quad \ge \alpha \lambda_n(a_1) + \beta \lambda_n(a_2).
\end{array}
$$
In the limit $n \to \infty$ this implies that the function $\lambda(a)$ is
concave.

Finally, lemma~\ref{l:updown} implies that $-M \le (1/n)L(0,y;n,x) -
L_0((x-y)/n) \le -m$. Hence,
$$
   m + a{x-y\over n} - L_0\left({x-y\over n}\right)
   \le a{x-y\over n} - {1\over n}L(0,y;n,x)
   \le M + a{x-y\over n} - L_0\left({x-y\over n}\right).
$$
Taking $\max$ over $(y,x) \in \rset^2$, using~(\ref{defHn}), and
denoting $(x-y)/n$ by $v$, we obtain
$$
   m + \max_{v \in \rset}\,(av - L_0(v)) \le
   -\lambda_n(a) \le
   M + \max_{v \in \rset}\,(av - L_0(v)).
$$
But by a well-known formula for the Legendre transform
$\max_{v \in \rset}\,(av - L_0(v)) = H_0(a)$. Thus we
obtain~(\ref{estHH0rep}) in the limit $n \to \infty$. \hfill$\square$

\begin{thm}
Suppose $a \in \rset$ and $s_0(x)$ is a continuous and periodic function;
then there exists
\begin{equation}
   \liminf_{n \to \infty}\,(s_n(x;a,s_0) - n\lambda(a)) = s^a(x)
\label{saliminf}
\end{equation}
and $s^a(x)$ satisfies functional equation~(\ref{functeq}).
\label{t:I}
\end{thm}

\textbf{Proof.} Using definitions of $s_n(x;a,s_0)$ and $\lambda_n(a)$, we
get
$$
   \min_{y \in \rset}\,s_0(y) + n\lambda_n(a)
   \le s_n(x;a,s_0)
   \le \max_{y \in \rset}\,s_0(y) + n\lambda_n(a).
$$
Subtracting $n\lambda(a)$ and using~(\ref{HnHa}), we obtain
$$
   \min_{y \in \rset}\,s_0(y) - C^*(a)
   \le s_n(x;a,s_0) - n\lambda(a)
   \le \max_{y \in \rset}\,s_0(y) + C^*(a).
$$
Let
$$
   \bar s_n(x) = \inf_{m \ge n}\,(s_m(x;a,s_0) - m\lambda(a)),\quad
   n = 1,2,\ldots.
$$
For any $x \in \rset$ the sequence $\{\bar s_n(x)\}$ is nondecreasing; we
see that it is bounded. Further, the corollary to lemma~\ref{l:LanLip}
implies that all functions $\bar s_n(x)$ are Lipschitzian with the constant
$C^*(a)$; hence, this sequence is equicontinuous. It follows that there
exists
$$
   \lim_{n \to \infty} \bar s_n(x)
   = \liminf_{n \to \infty}\,(s_n(x;a,s_0) - n\lambda(a))
   = s^a(x)
$$
and $s^a(x)$ is periodic and continuous.

Let us check that $s^a(x)$ satisfies functional equation~(\ref{functeq}). We have
$$
\begin{array}{l}
   \ds\min_{y \in \rset}\,(\bar s_n(y) + L^a_1(y,x)) - \lambda(a) = \\
   \ds\quad = \min_{y \in \rset}\,(\inf_{m \ge n}\,(\min_{z \in \rset}\,
   (s_0(z) + L^a_m(z,y) + L^a_1(y,x) - (m+1)\lambda(a)))).
\end{array}
$$
But it follows from the definition of $\lambda_n(a)$ and lemma~\ref{l:Ha}
that $L^a_m(z,y) - m\lambda(a) \ge -C^*(a)$ for all $m = 1,2,\ldots$
Therefore we can take minimum over $y \in \rset$ before infimum over $m \ge
n$ and obtain
$$
   \min_{y \in \rset}\,(\bar s_n(y) + L^a_1(y,x)) - \lambda(a) =
   \bar s_{n+1}(x).
$$
Taking into account~(\ref{defLan}) and passing to the limit $n \to \infty$,
we see that $s^a(x)$ satisfies functional equation~(\ref{functeq}). \hfill$\square$

\section{Existence of two-sided minimizers}

Suppose $a \in \rset$, $s^a(x)$ is a continuous periodic function
satisfying functional equation~(\ref{functeq}). Denote $L_a(y,x) = L(0,y;1,x) + a(y-x) -
\lambda(a)$. Let $Y^a(x)$ be the many-valued map of $\rset$ to $\rset$
defined as
\begin{equation}
   Y^a(x) = \arg \min_{y \in \rset}\,(s^a(y) + L_a(y,x)).
\label{defYa}
\end{equation}
It is readily seen that if $x$, $y \in \rset$, then $y \in Y^a(x)$ iff
\begin{equation}
   s^a(y) + L_a(y,x) = s^a(x)
   = \min_{z \in \rset}\,(s^a(z) + L_a(z,x)).
\label{ncc}
\end{equation}

For any two sets $X \subset \rset$, $Y \subset \rset$, denote by $X + Y$
the set $\{\, z \in \rset \mid z = x + y, x \in X, y \in Y \,\}$. If $Y =
\{y\}$, we denote $X + \{y\}$ by $X + y$.

\begin{lm}
For all $x \in \rset$
\begin{equation}
   Y^a(x+1) = Y^a(x) + 1.
\label{Yper}
\end{equation}
If there exists $x_0 \in \rset$ such that $x \in Y^a(x_0)$, then the set
$Y^a(x)$ contains only one point. If $x_1$, $x_2$, $y_1$, $y_2 \in \rset$,
$x_1 < x_2$, $y_1 \in Y^a(x_1)$, and $y_2 \in Y^a(x_2)$, then $y_1 < y_2$.
\label{l:Yper}
\end{lm}
\textsc{Proof.} Equation~(\ref{Yper}) follows from periodicity of $s^a(x)$
and diagonal periodicity of $L_a(y,x)$ (lemma~\ref{l:diagper}).

Suppose $x_1 < x_2$ and $y_i \in Y^a(x_i)$, $i = 1,2$. By~(\ref{ncc}), it
follows that
$$
\begin{array}{l}
   s^a(y_1) + L(0,y_1;1,x_1) + ay_1 \le s^a(y_2) + L(0,y_2;1,x_1) + ay_2, \\
   s^a(y_2) + L(0,y_2;1,x_2) + ay_2 \le s^a(y_1) + L(0,y_1;1,x_2) + ay_1,
\end{array}
$$
that is
$$
   L(0,y_1;1,x_1) + L(0,y_2;1,x_2) \le L(0,y_2;1,x_1) + L(0,y_1;1,x_2).
$$
Thus it follows from the corollary to lemma~\ref{l:Bangert} that $y_1
\le y_2$.

Let us check that $y_1 < y_2$. Assume the converse; let $y = y_1 = y_2$.
Take any $y_0 \in Y^a(y)$. By~(\ref{ncc}), it follows that
$$
   s^a(x_i) = s^a(y_0) + L(0,y_0;1,y) + L(1,y;2,x_i) + a(y_0 - x_i)
   - 2\lambda(a),
   \quad i = 1,2.
$$
On the other hand, using~(\ref{functeq}), we get
$$
\begin{array}{l}
   \ds s^a(x_i) = \min_{z \in \rset} \min_{t \in \rset}\,
   (s^a(t) + L(0,t;1,z) + L(1,z;2,x_i) + a(t - x_i)
   - 2\lambda(a)) \le \\
   \ds \qquad \le \min_{z \in \rset}\,
   (s^a(y_0) + L(0,y_0;1,z) + L(1,z;2,x_i) + a(y_0 - x_i)
   - 2\lambda(a)), \\
   \qquad i = 1,2.
\end{array}
$$
Hence it follows from lemma~\ref{l:Bangert} that $x_1 = x_2$; this
contradiction proves that $y_1 < y_2$.

Finally, suppose $x \in Y^a(x_0)$ for some $x_0 \in \rset$ and $y_1$, $y_2
\in Y^a(x)$. By a similar argument, we see that $y_1 = y_2$. Thus
$Y^a(x)$ consists of a single point. \hfill$\square$

\begin{lm}
Suppose $M_0 = \rset$, $M_n = Y^a(M_{n-1})$, $n = 1,2,\ldots;$ then $M_0
\supset M_1 \supset M_2 \supset \ldots$ and there exists a nonempty closed
set $M^a$ such that (i) $M^a = \bigcap_{n = 0}^\infty M_n$; (ii) for any
open $U \subset \rset$ such that $M^a \subset U + \zset$ there exists $N >
0$ with the following property: if the sequence $\{y_n\}$ is such that
$y_{n+1} \in Y^a(y_n)$, $n = 0,1,\ldots,$ then $y_n \in U + \zset$ for all
$n \ge N$; (iii) the restriction $Y^a|_{M^a}$ of $Y^a$ to $M^a$ is a
one-to-one continuous mapping with continuous inverse.
\label{l:Ma}
\end{lm}
\textsc{Proof.} Evidently, all $M_n \neq \varnothing$ and $M_0 = \rset$
contains $M_1$. Suppose $M_n \subset M_{n-1}$ for some $n \ge 1$, $x \in
M_n$; then $Y^a(x) \subset Y^a(M_n) \subset Y^a(M_{n-1}) = M_n$.  Hence
$M_{n+1} = Y^a(M_n) \subset M_n$.

Let us show that all $M_n$, $n = 1,2,\ldots,$ are closed. Note that $M_0 =
\rset$ is closed. Suppose $M_n$ is closed and the sequence $y_k \in
M_{n+1}$, $k = 1,2,\ldots,$ converges to a limit $\bar y \in \rset$. It
follows from lemma~\ref{l:Yper} that for any $k$ there exists a unique $x_k
= (Y^a)^{-1}(y_k) \in M_n$ and all $x_k$ are contained in a bounded
interval.  Thus there exists a subsequence $\{x_{k_l}\}$ that converges to
a limit $\bar x \in M_n$. Using~(\ref{ncc}), we get
$$
   s^a(y_{k_l}) + L_a(y_{k_l},x_{k_l}) = s^a(x_{k_l}).
$$
Since the finctions $s^a(x)$ and $L_a(y,x)$ are continuous, we obtain
$$
   s^a(\bar y) + L_a(\bar y,\bar x) = s^a(\bar x).
$$
Now it follows from~(\ref{ncc}) that $\bar y \in Y^a(\bar x)$. Hence $\bar
y \in Y^a(M_n) = M_{n+1}$, that is $M_{n+1}$ is closed.

Consider the topology $\mathcal{T}$ on $\rset$ such that $V \in \mathcal{T}$
iff $V = U + \zset$, where $U$ is open in the usual sense. Note that
$\rset$ is compact in this topology. It follows from~(\ref{Yper}) that
complements $U_n$ of sets $M_n$ are open in the topology $\mathcal{T}$.
We see that $U_n \subset U_{n+1}$.

Clearly, the set $M^a = \bigcap_{n=0}^\infty M_n$ is closed. Let us check
that it is nonempty. Assume the converse; hence sets $U_n$, $n =
1,2,\ldots$ cover all $\rset$. Since $\rset$ is compact in the topology
$\mathcal{T}$, we see that there exists $N > 0$ such that $\rset =
\bigcup_{n = 0}^N U_n = U_N$. Thus $M_N = \varnothing$; this contradiction
proves that $M^a$ is nonempty.

Now let $V \in \mathcal{T}$ be such that $M^a \subset V$. Arguing as above,
we see that there exists $N > 0$ such that $\rset = V \cup U_N$. On the
other hand, for any sequence $\{y_n\}$ such that $y_{n+1} \in Y^a(y_n)$, $n
= 0,1,\ldots,$ it follows that $y_n \in M_n$. Thus for all $n \ge N$ we
obtain $y_n \in V$.

Denote by $Y^a|_{M_n}$:~$M_n \to M_{n+1}$ the restriction of $Y^a$ to
$M_n$, $n = 1,2,\ldots$ It follows from lemma~\ref{l:Yper} that
$Y^a|_{M_n}$ is single-valued, strictly increasing as a function $\rset
\to \rset$, and bijective.  We claim that it is a homeomorphism. Indeed,
let sequences $\{x_k\} \subset M_n$ and $\{y_k\} \subset M_{n+1}$ be such
that $Y^a(x_k) = y_k$ for all $k$. If $y_k$ converge to $\bar y$, then it
follows from strict monotonicity of $Y^a|_{M_n}$ that $x_k$ converge to a
unique $\bar x$; hence the map $Y^a|_{M_n}$ has a continuous inverse.
Further, if $x_k$ converge to $\bar x$, then the set $\{y_k\}$ is bounded.
Let $\bar y_1$, $\bar y_2$, $y_1 \neq y_2$, be two limit points of
$\{y_k\}$; then we see that $\{\bar y_1,\bar y_2\} \subset Y^a(\bar x)$,
since $s^a(x)$ and $L_a(y,x)$ are continuous. But the map $Y^a|_{M_n}$ is
single-valued; this contradiction proves that it is continuous.

By the above $M^a$ is a closed subset of $M_n$ for all $n = 1,2\ldots;$
in addition, $Y^a(M^a) = M^a$. It follows that the restriction of
$Y^a|_{M_n}$ to $M^a$ is a homeomorphism. \hfill$\square$

We say that $M^a$ is the \emph{invariant set} of the many-valued map
$Y^a(x)$.

Combining the previous lemma with the observation that sequence $\{x_n\}$
that satisfies~(\ref{inclusion}) is $L$-minimal, we obtain the following
\begin{thm}
Suppose $a \in \rset$, $s^a(x)$ is a continuous periodic function
satisfying functional equation~(\ref{functeq}), $Y^a(x)$ is the corresponding many-valued map,
$M^a$ is its invariant set, and $x \in M^a$. Then the two-sided sequence
$x_k = (Y^a)^{-k}(x)$, $k \in \zset$, is $L$-minimal.
\label{t:Iminimizer}
\end{thm}

\section{Uniqueness of an eigenfunction}

To give the uniqueness condition for eigenfunction of the Bellman operator
we need the following results of Aubry-Mather theory:
\begin{lm}
Suppose a sequence $\{x_n\}$, $n \in \zset$, is $L$-minimal; then there
exists
\begin{equation}
   \lim_{n \to \pm\infty} {x_n - x_0\over n} = \omega(\{x_n\}) \in \rset.
\label{defomega}
\end{equation}
If $\omega = \omega(\{x_n\})$ is irrational, then there exists a function
$\phi_\omega(t)$:~$\rset \to \rset$ with the following properties:
(i) it is continuous on the right and $\phi_\omega(t_1) < \phi_\omega(t_2)$
iff $t_1 < t_2$;
(ii) $\phi_\omega(t + 1) = \phi_\omega(t) + 1$ for all $t \in \rset$;
(iii) if $x_0 = \phi_\omega(t_0 \pm 0)$ then the sequence $\{x_n\}$ defined
by $x_n = \phi_\omega(t_0 + n\omega \pm 0)$ is $L$-minimal;
(iv) for any neighborhood $V \in \mathcal{T}$ of the closure of the image
$\phi_\omega(\rset)$ and any $L$-minimal sequence $\{x_n\}$, $x_n \in
V$ as soon as $|n|$ is large enough.
\label{l:MF}
\end{lm}
For the proof see, e.g.,~\cite[sections~9,~11, and~12]{MF}. The number
$\omega(\{x_k\})$ is called a \emph{rotation number} of the $L$-minimal
sequence $\{x_k\}$.

\begin{thm}
Let $s^a(x)$ be the continuous periodic function
satisfying functional equation~(\ref{functeq}), $Y^a(x)$ be the corresponding many-valued map
and $M^a$ be its invariant set. Then the value of $\omega$ for any
$L$-minimal sequence of the form $x_n = {(Y^a)}^{-k}(x)$ depends only on
the parameter $a$ but not on the choice of $s^a(x)$; furthermore, if
$\omega(a)$ is irrational, then $s^a(x)$ is determined uniquely up to an
additive constant.
\label{t:II}
\end{thm}

\textsc{Proof.} Let $Y^a(x)$ be a many-valued map corresponding to a
countinuous periodic function $s^a(x)$ satisfying functional
equation~(\ref{functeq}) and $M^a$ be its invariant set. The corresponding
rotation number $\omega$ is determined uniquely. Indeed, let $x_k =
(Y^a)^{-k}(x)$ and $y_k = (Y^a)^{-k}(y)$, $k \in \zset$, for some $x$,~$y
\in M^a$; we are going to demonstrate that $\omega(\{x_k\}) =
\omega(\{y_k\})$. Take $p \in \zset$ such that $x \le y + p \le x + 1$. By
lemma~\ref{l:Yper}, it follows that $(Y^a)^{-k}(x) \le (Y^a)^{-k}(y + p)
\le (Y^a)^{-k}(x + 1)$ for all $k \in \zset$.  Using~(\ref{defomega}), we
obtain $\omega(\{x_k\}) = \omega(\{y_k\})$.

A subset $M \subset M^a$ is called \emph{a minimal set} of the
(single-valued) map $Y^a|_{M^a}$ if $M$ is closed, $Y^a(M) = M$, and $M$
contains no other nonempty subset with the same properties. Let $x \in
M^a$; the sequence $\{(Y^a)^k(x)\}$, $k = 1,2,\ldots,$ is called \emph{the
orbit} of the point $x$. It can be proved that a set $M \subset M^a$ is
minimal if and only if the orbit of any point $x \in M$ is dense in $M$ in
topology $\mathcal{T}$ defined in the proof of lemma~\ref{l:Ma}.

The set $M^a$ is compact in topology $\mathcal{T}$; using~(\ref{Yper}), we
see that $Y^a$ is a homeomorphism of $M^a$ in this topology. By the
well-known theorem of Birkhoff~\cite{B}, it follows that $M^a$ contains
at least one minimal set $M$ of the map $Y^a$.

Suppose $s^a_1(x)$ is another continuous periodic function satisfying
functional equation~(\ref{functeq}); then for all $x$, $y \in \rset$ such
that $y \in Y^a(x)$
$$
    s^a(x) - s^a_1(x) \ge s^a(y) - s^a_1(y).
$$
Indeed, it follows from~(\ref{ncc}) that $s^a(x) = s^a(y) + L_a(y,x)$.
Using~(\ref{defsn}), we obtain
$$
\begin{array}{rl}
   s^a(x) - s^a_1(x)
   &\ds\!\!\!\!= s^a(y) + L_a(y,x) - s^a_1(x) = \\
   &\ds\!\!\!\!= s^a_1(y) + L_a(y,x) - s^a_1(x) + s^a(y) - s^a_1(y) \ge \\
   &\ds\!\!\!\!\ge s^a(y) - s^a_1(y).
\end{array}
$$

Clearly, $s^a(x)$ and $s^a_1(x)$ are continuous in topology $\mathcal{T}$.
Let $x_0 \in M$ be a point where $s^a(x) - s^a_1(x)$ attains its minimal
value in $M$ and $\{x_k\}$, $k = 1,2,\ldots,$ be its orbit. By the above,
$$
   s^a(x_0) - s^a_1(x_0) \le s^a(x_k) - s^a_1(x_k)
   \le s^a(x_0) - s^a_1(x_0).
$$
Since the closure of $\{x_k\}$ is the set $M$, we see that on $M$ $s^a(x) -
s^a_1(x)$ is constant. Thus it follows from~(\ref{ncc}) that on the set $M$
$Y^a(x)$ coincides with the map $Y^a_1(x)$ corresponding to $s^a_1(x)$. In
particular, the rotation number $\omega$ depends only on $a$. We see also
that if $M$ is a minimal set of the map corresponding to some $s^a(x)$
satisfying functional equation~(\ref{functeq}), then it is a minimal set of
any other continuous periodic function satisfying functional
equation~(\ref{functeq}) with the same $a$.

Now suppose $\omega = \omega(a)$ is irrational. It follows from
lemma~\ref{l:MF}~(iv) that the closure $\bar M$ of the set
$\phi_\omega(\rset)$ contains limit points with respect to topology
$\mathcal{T}$ of the closed set $M^a$, so $\bar M \cap M^a$ is nonempty.
Now if $x_0 \in M^a$ and there exists such $t_0$ that $x_0 =
\phi_\omega(t_0 \pm 0) \in \bar M$, then by lemma~\ref{l:MF}~(iii) the
closure of the orbit of the point $x_0$ with respect to topology
$\mathcal{T}$ coincides with $\bar M$, so $\bar M \subseteq M^a$. Thus
statement~(iv) of lemma~\ref{l:MF} means that $\bar M$ is the only minimal
invariant subset of $M^a$ not depending on the particular choice of an
eigenfunction $s^a(x)$.

Finally let us prove that if $\omega = \omega(a)$ is irrational, then
$s^a(x) - s^a_1(x)$ is constant for all $x \in \rset$. Without loss of
generality it can be assumed that $s^a(x) - s^a_1(x) = 0$ on $\bar M$. Let
$x_0 \in \rset$ be a point where $s^a(x) - s^a_1(x)$ attains its minimum
in $M^a$ and $\{x_k\}$, $k = 1,2,\ldots,$ be its orbit. Arguing as above,
we see that $s^a(x_k) - s^a_1(x_k) = s^a(x_0) - s^a_1(x_0)$ for all $k
\ge 1$. On the other hand, it follows from lemma~\ref{l:MF}~(iv) that for
any neighborhood $V \in \mathcal{T}$ of the set $\bar M$ there exists $N >
0$ such that $x_k \in V$ for all $k > N$. Using continuity of the functions
$s^a(x)$ and $s^a_1(x)$, we see that $s^a(x_0) - s^a_1(x_0) = 0$. By
definition of $x_0$, it follows that $s^a(x) \ge s^a_1(x)$ for all $x \in
M^a$. Likewise, $s^a(x) \le s^a_1(x)$; thus $s^a(x) = s^a_1(x)$ on $M^a$. A
similar argument based on lemma~\ref{l:Ma}~(ii) shows that $s^a(x) =
s^a_1(x)$ for all $x \in \rset$.
\hfill$\square$

\eject

\end{document}